\documentclass[prd,aps,twocolumn,a4paper,floatfix]{revtex4}

\def\p{\partial}
\def\Lie{{\cal L}}

\def\ul{\underline}
\def\perpn{\perp\!}
\def\perpN{{}^{\textrm{\tiny{(N)}}}\!\!\!\perp}

\def\gamN{{}^{\textrm{\tiny{(N)}}}\!\gamma}

\def\DN{{}^{\textrm{\tiny{(N)}}}\!D}

\def\KN{{}^{\textrm{\tiny{(N)}}}\!K}

\def\EN{{}^{\textrm{\tiny{(N)}}}\!E}
\def\BN{{}^{\textrm{\tiny{(N)}}}\!B}

\newcommand{\sLie}{\mathcal{L}\mkern-9mu /}
\newcommand{\sD}{D\mkern-12mu /}
\newcommand{\sG}{\Gamma\mkern-12mu /}
\newcommand{\sP}{\p\mkern-10mu /}
\newcommand{\sR}{R\mkern-12mu /}

\usepackage{graphicx,psfrag}
\usepackage{amsmath,amsfonts,amssymb,amsthm}
\usepackage[mathscr]{euscript}

\usepackage{color}
\usepackage{float}
\usepackage{ulem}

\usepackage{comment}
\usepackage{bbm}

\begin{document}


\title{Dual Foliation Formulations of General Relativity}

\author{David Hilditch}
\affiliation{Theoretical Physics Institute, University of 
Jena, 07743 Jena, Germany}

\date{September 8, 2015}

\begin{abstract}
A dual foliation treatment of General Relativity is presented. The basic 
idea of the construction is to consider two foliations of a spacetime 
by spacelike hypersurfaces and relate the two geometries. The 
treatment is expected to be useful in various situations, and in 
particular whenever one would like to compare objects represented in 
different coordinates. Potential examples include the construction of 
initial data and the study of trapped tubes. It is common for studies in 
mathematical relativity to employ a double-null gauge. In such studies 
local well-posedness is treated by referring back, for example, to the 
generalized harmonic formulation, global properties of solutions being 
treated in a separate formalism. As a first application of the dual 
foliation formulation we find that one can in fact obtain local well-posedness 
in the double-null coordinates {\it directly}, which should allow their 
use in numerical relativity with standard methods. With due care it is 
expected that practically any coordinates can be used with this approach. 
\end{abstract}

\maketitle

\section{Introduction}\label{section:Introduction}

For their consideration as an initial value problem the field 
equations of General Relativity are typically split into a set 
of evolution  and constraint equations. This is done by introducing 
coordinates~$x^\mu=(t,x^i)$. The level sets of the time coordinate~$t$ 
are taken to be spacelike hypersurfaces which foliate the spacetime. 
The unit normal to the hypersurfaces is used to~$3+1$ decompose the 
field equations in the natural way. This results in the vacuum 
field equations in the textbook form~\cite{Gou07,Alc08,BauSha10},
\begin{align}
\p_t\gamma_{ij}&=-2\alpha K_{ij}+\Lie_\beta\gamma_{ij}\,,\nonumber\\
\p_tK_{ij}&=-D_iD_j\alpha+\alpha[R_{ij}-2K^k{}_iK_{jk}+KK_{ij}]
+\Lie_\beta K_{ij}\,,\nonumber\\
H&=R-K_{ij}K^{ij}+K^2=0\,,\nonumber\\
M_i&=D^jK_{ij}-D_iK=0\,.\label{eqn:ADM_eqns}
\end{align} 
Given two sets of observers, one associated with~$x^\mu$, another 
with coordinates~$X^{\ul{\mu}}=(T,X^{\ul{i}})$, what is the relationship 
between the solutions as expressed in each {\it in the~$3+1$ picture?} 
Unfortunately a clear presentation of the resulting formalism is not 
readily available, despite being straightforwardly obtained by  
space-time decomposition of the four-dimensional 
Jacobians~$J^{\ul{\mu}}{}_\mu=\p X^{\ul{\mu}}/\p x^{\mu}$. The first aim 
of this work is to give just such a presentation, which can be found 
in section~\ref{section:GR}. This dual foliation approach will be useful in 
numerical relativity, where one expects it will help in the construction 
of initial data and in the comparison of solutions constructed with 
different choices of lapse~$\alpha$ and shift~$\beta^i$.

Amongst the most powerful machinery in mathematical relativity 
is that of the double-null coordinates. With this choice the field 
equations exhibit a particular structure that allows the demonstration
of the stability of Minkowski spacetime~\cite{KlaNic03}, and that a 
certain special class of vacuum initial data will collapse to form 
blackholes~\cite{Chr08}. One would thus like to use these coordinates in 
numerical relativity, preferably with standard methods. A number of applications 
present themselves; the conjectured instability of Cauchy 
horizons~\cite{SimPen73}, the propagation of weak-null 
singularities~\cite{Luk13}, and the {\it critical formation of 
blackholes}~\cite{GunGar07}. Unfortunately from this perspective the proofs 
of these impressive results employ a different formalism for long-term 
results and local existence. This is a serious bugbear because, as painful 
experience has taught, a necessary condition for any numerical method to 
converge is that the PDE problem is locally well-posed. Therefore the second 
aim of this work is to find such a formulation. In section~\ref{section:DN} 
this is attempted in a direct way. The normal approach is to modify the 
equations by introducing new constraints, coupled in a particular way to the 
gauge choice, and insodoing uncover, say, a strongly hyperbolic formulation. But 
we find in a pure gauge analysis that the standard form of the double-null 
coordinates are only weakly hyperbolic, and so can not be used in this 
way~\cite{KhoNov02,HilRic13,Hil13}. With appropriate alterations, there 
may be such a straightforward formulation, but because a preferred direction 
is singled out the construction will in any case be complicated. We thus 
abandon the search.

In section~\ref{section:_3+1_GHG11} we summarize the first order 
reduction~\cite{LinSchKid05} of the generalized harmonic gauge (GHG) 
formulation~\cite{Bru52,Fri85,Gar01} employed in the numerical relativity codes 
SpEC~\cite{SpEC} and~\texttt{bamps}~\cite{Bru11,HilWeyBru15,BugDieBer15}. We use the dual 
foliation formalism to derive equations of motion for the Jacobian that maps 
from generalized harmonic to double-null coordinates. These equations are 
minimally coupled to the field equations, and so their hyperbolicity may be 
treated easily. Indeed the Jacobians satisfy a set of nonlinear advection 
equations, and so are hyperbolic. We may consider the full set of fields to 
be solved for as the GHG system with the Jacobians tacked 
on. We can subsequently change independent variables 
from~$x^\mu\to X^{\ul{\mu}}$, with~$X^{\ul{\mu}}$ the double-null coordinates. 
The punchline is that since the system has at most first derivatives, we can 
do so {\it without generating any derivatives of the evolved Jacobians}. 
Therefore the PDE properties of the system are unaffected and we end up 
with a formulation that is symmetric hyperbolic {\it directly} in the 
double-null coordinates. Weaker notions of hyperbolicity are also 
considered. Concluding remarks are collected in 
section~\ref{section:Conclusion}. Geometric units are used 
throughout.

\section{The Dual Foliation Formalism}
\label{section:GR}

In this section we work in the intersection of two coordinate 
patches~$x^{\mu}=(t,x^i)$ and~$X^{\ul{\mu}}=(T,X^{\ul{i}})$, related 
by the Jacobian~$J^{\ul{\mu}}{}_\mu=\p X^{\ul{\mu}}/\p x^{\mu}$. The two 
time coordinates define distinct foliations of the spacetime, and 
with them different notions of spacelike tensors, intrinsic and 
extrinsic curvatures. These quantities are related in the natural 
way with a~$3+1$ split of the Jacobian. Consequently the form of 
the gravitational field equations in each foliation is related. 
Throughout latin indices~$a,b,c,d,e$ will be abstract. Greek 
indices stand for those in coordinates~$x^\mu$, or if underlined
in~$X^{\ul{\mu}}$. Similarly latin indices~$i,j,k,l,m,p$ stand
for spatial components in~$x^\mu$, and when underlined for spatial 
components in~$X^{\ul{\mu}}$. Indices~$n$ and~$v$ denote contraction
in that slot with the vectors~$n^a$ and~$v^a$ respectively. The 
summation convention is always employed.

\subsection{Coordinate freedom}
\label{section:Pure_Gauge}

\paragraph*{Coordinate change under a $3+1$ decomposition:}
Consider two sets of coordinates~$x^{\mu}$ and~$X^{\ul{\mu}}$ defined
in the same region of spacetime. Each of the two time 
coordinates~$t$ and~$T$ naturally defines a foliation of the spacetime 
which we will refer to as the lower case and upper case foliations 
respectively. In the lower case foliation we define the standard 
lapse, normal vector, time vector, projection operator, and shift 
vector,
\begin{align}
\alpha&=(-\nabla_at\,\nabla^at)^{-\frac{1}{2}}\,,\quad 
&n^a=-\alpha\nabla^at\,,\nonumber\\
&t^a\nabla_at\equiv 1\,,
&\perpn\!{}^a{}_b=\delta^a{}_b+n^an_b\,,\nonumber\\
\beta_a&=\perpn^b{}_at_b\,,&
\beta^i=-\alpha n^a\nabla_ax^i\,.
\label{eqn:3+1_x}
\end{align}
With both indices downstairs the projection 
operator~$\perp^a\!\!{}_b$ becomes the natural induced 
metric~$\gamma_{ab}$ on the lower case foliation. The covariant 
derivative associated with~$\gamma_{ab}$ is denoted by~$D$ with 
connection~$\Gamma$. The same 
definitions are made in the upper case foliation, 
\begin{align}
A&=(-\nabla_aT\,\nabla^aT)^{-\frac{1}{2}}\,,\quad 
&N^a=-A\nabla^aT\,,\nonumber\\
&T^a\nabla_aT\equiv 1\,,
&\perpN^a\!\!{}_b=\delta^a{}_b+N^aN_b\,,\nonumber\\
B_a&=\perpN^b\!\!{}_aT_b\,, &
B^{\ul{i}}=-A N^a\nabla_aX^{\ul{i}}\,.\label{eqn:3+1_X}
\end{align}
The covariant derivative associated with~$\gamN_{ab}$ is 
denoted by~$\DN$ with connection~${}^{\textrm{\tiny{(N)}}}\Gamma$. 

\begin{table}[t]
  \centering
  \begin{ruledtabular}
  \caption{\label{tab:Metrics} 
  A summary of the definitions of the various metrics, time reduction
  variables the relationship between them. The fourth column gives  
  the object used as a time reduction variable when employing the given 
  metric. The final column gives states the equation numbers relating the 
  curvature of the given metric with that of the others. `GCM' stands for 
  the Gauss-Codazzi-Mainardi equations.}
  \begin{tabular}{cccccc}
    Metric & Connection & Defn. & Time der. & Curv. & Note\\
    \hline
    \hline
   $g_{ab}$ & $\nabla,{}^{\tiny{(4)}}\Gamma$ &   &  & ${}^{\tiny{(4)}}R_{ab}$ & \\
   $\gamma_{ab}$ & $D,\Gamma$ & $\perp g_{ab}$ & $K_{ab}$  & $R_{ab}$ & `GCM'\\
   $\mathbbmss{g}_{ab}$ & $\mathbbmss{D},\mathbbmss{G}$ & $\perp\gamN_{ab}$ 
 & $\mathbbmss{K}_{ab}$ & $\mathbbmss{R}_{ab}$ & \eqref{eqn:BoostRvsR} \\
   $\gamN_{ab}$ & $\DN,{}^{\textrm{\tiny{(N)}}}\Gamma$ & $\perpN g_{ab}$ & $\KN_{ab}$ 
 & ${}^{\textrm{\tiny{(N)}}}\!R_{ab}$ & \eqref{eqn:BoostRvsRN} \\
  \end{tabular}
  \end{ruledtabular}
\end{table}

\paragraph*{The Lorentz factor and boost vector:} The unit 
normal vectors of the upper and lower case foliations are 
related by 
\label{eqn:n->N}
\begin{align}
N^a&=W(n^a+v^a),
\end{align}
where we have defined the Lorentz factor~$W$ and lower case 
boost vector~$v_a$,
\begin{align}
W& = -(N^an_a)\,,&\quad
v_a=\frac{1}{W}\perp^b\!\!{}_aN_b\,.
\end{align}
Since the normal vectors have unit magnitude the 
Lorentz factor and boost vector satisfy,
\begin{align}
W&=\frac{1}{\sqrt{1-v_iv^i}}\,,&\qquad W\geq1>\gamma^{ij}v_iv_j
\equiv v^2\,.
\end{align}
This is simply the requirement that the upper and lower case normal 
observers travel subluminally relative to one another. Observe that we 
also have,
\begin{align}
n^a=W(N^a+V^a)\,,
\end{align}
with the upper case boost vector,
\begin{align}
V_a&=\frac{1}{W}\perpN^b\!\!{}_an_b=(W^{-1}-W)n_a-W v_a\,,
\end{align}
so that there is a natural reciprocity between the coordinate
systems as expected. We thus also obtain,
\begin{align}
W=\frac{1}{\sqrt{1-V_{\ul{i}}V^{\ul{i}}}}\,.
\end{align}
Provided the spatial boost vector, a vector~$S_a$ which is spacelike in 
the upper case foliation, $S_aN^a=0$, can be reconstructed from its 
projection into the lower case foliation with,
\begin{align}
S_a=(\perpn\!S)_b(\delta^b{}_a+v^bn_a),\label{eqn:full_from_proj}
\end{align}
and obviously a similar result holds for all tensor valences. Therefore 
we may restrict our attention of upper case spacelike tensors to their 
projections into the lower case foliation, and thus look only at the 
spatial components in the lower case tensor basis. 

\paragraph*{Decomposition of the Jacobian~$J^{\ul{\mu}}{}_{\mu}$:}
Upon~$3+1$ decomposition the 
Jacobian~$J^{\ul{\mu}}{}_\mu\equiv\p X^{\ul{\mu}}/\p x^{\mu}$ can be 
written,
\begin{align}
n^\alpha J^{\ul{\alpha}}{}_{\alpha}N_{\ul{\alpha}}&=-W\,,
\quad\quad
n^\alpha J^{\ul{i}}{}_{\alpha} 
\equiv\pi^{\ul{i}}\,, \nonumber\\
J^{\ul{\alpha}}{}_iN_{\ul{\alpha}}&=Wv_i\,,
\quad\quad
J^{\ul{i}}{}_i\equiv \phi^{\ul{i}}{}_i\,.
\end{align}
The components~$\pi^{\ul{i}}$ are given in terms of the upper 
case lapse, shift and boost vectors by,
\begin{align}
\pi^{\ul{i}}=WV^{\ul{i}}-WA^{-1}B^{\ul{i}}\,,
\label{eqn:Jac_Pi_Def}
\end{align}
although we rarely find that this is the most convenient form
for the expression. In matrix form we therefore have the 
representation,
\begin{align}
J&=\left(\begin{array}{cc}
 A^{-1}W(\alpha-\beta^iv_i) 
& \alpha\,\pi^{\ul{i}}+\beta^i\phi^{\ul{i}}{}_i \\
-A^{-1}Wv_i 
& \phi^{\ul{i}}{}_i
\end{array}\right)\,.
\end{align}
Note that by introducing the variables~$A^{-1}Wv_i$ and~$\phi^{\ul{i}}{}_i$ to 
replace first order spatial derivatives of the coordinates we have effectively 
introduced reduction constraints,
\begin{align}
D_{[i}(A^{-1}Wv_{j]})=0\,,\qquad D_{[i}\phi^{\ul{i}}{}_{j]}=0\,,
\label{eqn:Hyp_Cons}
\end{align}
which we will refer to as the hypersurface (orthogonality) 
constraints. Here and in what follows one must be careful to note that 
the upper case underlined index is to be treated as a simple label rather 
than an open slot when working in the lower case coordinates. It is 
straightforwardly checked that the transformation,
\begin{align}
J^{-1}&=\left(\begin{array}{cc}
\alpha^{-1}W(A-B^{\ul{i}}V_{\ul{i}}) & A\,\Pi^i+B^{\ul{i}}\Phi^i{}_{\ul{i}}\\
-\alpha^{-1}WV_{\ul{i}} & \Phi^i{}_{\ul{i}}
\end{array} \right)\,,\label{eqn:J_inv}
\end{align}
with the various auxiliary quantities defined in the natural 
way, is indeed the inverse transformation. 

\paragraph*{Time development of the Jacobian:} By the equality
of mixed partials we have the Hamilton-Jacobi like equations,
\begin{align}
\p_t(A^{-1}Wv_i)&=-D_i\big[\alpha (A^{-1}W)\big]+\Lie_\beta(A^{-1}Wv_i)\,,
\nonumber\\
\p_t\phi^{\ul{i}}{}_i&=D_i(\alpha\,\pi^{\ul{i}})+\Lie_\beta\phi^{\ul{i}}{}_i\,,
\label{eqn:Jac_dot}
\end{align}
for the components~$J^{\ul{T}}{}_i$ and~$J^{\ul{i}}{}_i$. These expressions 
hold regardless of the upper case coordinate choice, but the remaining 
four components can be determined only once a particular coordinate 
choice is known. Perhaps the simplest useful example is the generalized 
harmonic gauge choice~$\Box X^{\ul{\alpha}}=H^{\ul{\alpha}}$, which results in,
\begin{align}
\p_t(A^{-1}W)&=\alpha(A^{-1}W)(K+E)-D^i(\alpha A^{-1}Wv_i)\nonumber\\
&\quad+\Lie_\beta(A^{-1}W)\,,
\nonumber\\
\p_t\pi^{\ul{i}}&=D^j(\alpha\phi^{\ul{i}}{}_j)
+\alpha E^{\ul{i}}+\Lie_\beta\pi^{\ul{i}}\,,
\label{eqn:Jac_dot_GHG}
\end{align}
where the gauge source functions are decomposed as~$E=(A/W)\,H^{\ul{T}}$
and~$E^{\ul{i}}=-H^{\ul{i}}$. The lower case extrinsic curvature is defined by,
\begin{align}
K_{ab}&=-\perp^c\!\!{}_a\nabla_cn_b\,,
\end{align}
and likewise in the upper case foliation, except that as elsewhere 
we append a label~$N$. On a given spacetime with coordinates~$x^\mu$ 
the system~\eqref{eqn:Jac_dot},~\eqref{eqn:Jac_dot_GHG} can be viewed as 
a simple first order reduction of the four wave 
equations~$\Box X^{\ul{\mu}}=H^{\ul{\mu}}$. Other choices will be more 
conveniently expressed once the relationship between the two induced 
geometries are known.

\paragraph*{Equations of motion for projected upper case objects:} 
The equations of motion of the projection of upper case spacelike 
tensors~$S_a$ and~$S_{ab}$ projected into the lower case foliation 
are, 
\begin{align}
\p_tS_i&=\frac{\alpha}{W}\Lie_NS_i+\Lie_{(\beta-\alpha v)}S_i\,,\nonumber\\
\p_tS_{ij}&=\frac{\alpha}{W}\Lie_NS_{ij}+\Lie_{(\beta-\alpha v)}S_{ij}\,,
\end{align}
for vectors and symmetric tensors respectively. Similar 
expressions hold for arbitrary valences but will not be used in 
what follows. The projected upper case induced metric defined 
by~$\mathbbmss{g}_{ab}=\gamma^c{}_a\gamma^d{}_b\gamN_{cd}$ is,
\begin{align}
\mathbbmss{g}_{ij}=\gamN_{ij}&=\gamma_{ij}+W^2v_iv_j\,.\label{eqn:GmN->gmn}
\end{align}
Note that as the sum of a symmetric positive definite matrix and 
and semi-positive definite combination of the boost vector, the 
projected upper case metric is itself positive definite and thus invertible,
and can be considered a metric on leaves of the lower case foliation 
in its own right, if we so wish. When doing so we will refer to this 
object as the boost metric, and denote the covariant derivative 
by~$\mathbbmss{D}$ with connection~$\mathbbmss{G}$. The boost metric 
has inverse,
\begin{align}
(\mathbbmss{g}^{-1})^{ij}=\gamma^{ij}-v^iv^j\,,
\end{align}
by the Sherman-Morrison formula. We now aim to relate the geometry of 
the upper and lower case foliations in terms of the lower case normal, 
Lorentz factor and boost vector. A summary of the relationships between 
the four different metrics~$g_{ab},\gamma_{ab},\gamN_{ab}$ 
and~$\mathbbmss{g}_{ab}$ is given in Table~\ref{tab:Metrics}.

\paragraph*{Connections and curvatures:} The Christoffel 
symbol associated with the upper case induced metric 
is given by the standard expression,
\begin{align}
{}^{\textrm{\tiny{(N)}}}\Gamma^{\gamma}{}_{\mu\nu}&=
\perpN\!\!{}^{\tiny{(4)}}\Gamma^\gamma{}_{\mu\nu}\,,
\label{eqn:Upper_Connection}
\end{align}
which holds in arbitrary coordinates, and where here and in what 
follows we use the projection operator without indices to denote 
the projection on every open slot. The Christoffel symbol associated 
with the lower case induced metric is defined similarly. By the 
argument around equation~\eqref{eqn:full_from_proj} we need only compute 
the projection of the upper case Christoffel into the lower case 
foliation. Using~\eqref{eqn:GmN->gmn} 
and~\eqref{eqn:Upper_Connection} we find that,
\begin{align}
\label{eqn:Projected_Upper_Connection}
\!\!\perpn\!\!{}^{\textrm{\tiny{(N)}}}\Gamma^k{}_{ij}&=
\mathbbmss{g}^k{}_l\,\mathbbmss{g}^m{}_i\,\mathbbmss{g}^p{}_j\,\Gamma^l{}_{mp}
-2\,W^2\,\mathbbmss{g}^{kl}\,\mathbbmss{g}^m{}_{(i}v_{j)}K_{lm}
\nonumber\\
&+W^2v^k\,\mathbbmss{g}^m{}_i\,\mathbbmss{g}^p{}_j\,K_{mp}
-2\,W^4\,v^k\,\mathbbmss{g}^m{}_{(i}v_{j)}a_m\nonumber\\
&+W^4\mathbbmss{g}^{kl}\,v_iv_j\,a_l
+2W^2\alpha^{-1}\,\mathbbmss{g}^k{}_l\,\mathbbmss{g}^m{}_{(i}v_{j)}\p_m\beta^l
\nonumber\\
&+W^4\alpha^{-1}\,\mathbbmss{g}^k{}_l\,v_iv_j\,\p_n\beta^l
-W^6v^kv_iv_j\,\Lie_n\ln\alpha\,,
\end{align}
where here we use an index~$n$ to denote contraction with the lower case 
unit normal vector~$n^a$, and where the acceleration of lower case Eulerian
observers is~$a_i=D_i\ln\alpha$. The upper case induced Ricci tensor can be 
computed from,
\begin{align}
{}^{\textrm{\tiny{(N)}}}\!R_{\ul{\mu}\ul{\nu}}&=
\perpN\p_{\ul{\kappa}}{}_{}^{\textrm{\tiny{(N)}}}\Gamma^{\ul{\kappa}}{}_{\ul{\mu}\ul{\nu}}
-\perpN\p_{\ul{\mu}}{}_{}^{\textrm{\tiny{(N)}}}\Gamma^{\ul{\kappa}}{}_{\ul{\kappa}\ul{\nu}}
\nonumber\\
&+{}_{}^{\textrm{\tiny{(N)}}}\Gamma^{\ul{\kappa}}{}_{\ul{\mu}\ul{\nu}}
{}_{}^{\textrm{\tiny{(N)}}}\Gamma^{\ul{\delta}}{}_{\ul{\kappa}\ul{\delta}}
-{}_{}^{\textrm{\tiny{(N)}}}\Gamma^{\ul{\kappa}}{}_{\ul{\mu}\ul{\delta}}
{}_{}^{\textrm{\tiny{(N)}}}\Gamma^{\ul{\delta}}{}_{\ul{\nu}\ul{\kappa}}\,,
\end{align}
and likewise for the lower case curvature. The relationships between 
the upper and lower case connections above~\eqref{eqn:Upper_Connection} 
can be used to compute the relationship between the two Ricci curvatures 
by brute force, but it is more convenient to use the Gauss-Codazzi 
equations,
as described 
momentarily. The upper case extrinsic curvature projected into 
the lower case foliation is,
\begin{align}
\KN_{ij}&=WK_{ij}-D_{(i}Wv_{j)}-WA_{(i}v_{j)}\equiv W(\mathbbmss{K}_{ij}-A_{(i}v_{j)})\,,
\label{eqn:KNn}
\end{align}
where here we also define~$\mathbbmss{K}_{ij}$ which, for lack of a better 
name, we call the boost extrinsic curvature.  Note that we define the 
trace~$\mathbbmss{K}\equiv(\mathbbmss{g}^{-1})^{ij}\mathbbmss{K}_{ij}$ in a 
nonstandard way by using the inverse boost metric. The projected upper 
case acceleration is,
\begin{align}
\mathbbmss{A}_i\equiv A_i=D_i(\ln A)
+W^2v_i\Big(v^jD_j(\ln A)+\Lie_n(\ln A)\Big)\,.
\end{align}
It is most convenient to express the various equations in terms 
of~$\mathbbmss{A}_i$ rather than using this expression, as we wish 
to suppress the explicit appearance of a~$\Lie_n(\ln A)$ contribution. 

\paragraph*{Constraints under the coordinate change:} Comparing 
the Hamiltonian and momentum constraints in each foliation, we 
find that,
\begin{align}
{}^{\textrm{\tiny{(N)}}}H&=W^2H-2W^2M_v\,,\nonumber\\
{}^{\textrm{\tiny{(N)}}}M_i&=WM_i+2W^3M_vv_i-W^3H v_i\,.
\label{eqn:Constraints}
\end{align}
An index~$v$ denotes contraction with the boost vector~$v^a$.
It is thus obvious that the full set of constraints will be 
satisfied in the lower case foliation if and only if they are
satisfied in the upper case foliation. Expanding out 
the upper and lower case constraints and using projected upper 
case extrinsic curvature~\eqref{eqn:KNn} in combination 
with~\eqref{eqn:full_from_proj}, we easily find the relationship 
between the two spatial Ricci scalars. As it stands, 
equation~\eqref{eqn:Constraints} is really of a purely geometric 
nature and independent of the gravitational field equations, 
where we think of the symbols as mere shorthands according 
to~\eqref{eqn:ADM_eqns}, or the upper case foliation equivalent. 

\paragraph*{Electric and Magnetic decomposition of the Weyl 
tensor:} Especially in General Relativity the decomposition of 
the Weyl tensor~$W_{abcd}$ into two spatial tracefree tensors, 
Electric and Magnetic parts, 
\begin{align}
E_{ab}&=n^cn^dW_{acbd}\,,&\quad B_{ab}=n^cn^d\,{}^{\ast}W_{acbd}\,,
\label{eqn:E_B_defn}
\end{align}
has special significance in encoding the propagating degrees 
of freedom of the gravitational field. The dual Weyl tensor 
here is~${}^{\ast}W_{abcd}=\tfrac{1}{2}\epsilon^{ef}{}_{cd}W_{efab}$. 
Evidently this decomposition is foliation dependent, because of 
the presence of the normal vector~$n^a$. The relationship 
between the two decompositions is given by,
\begin{align}
\EN_{ij}&=(2W^2-1)E_{ij}-2W^2E_{v(i}v_{j)}+W^2E_{vv}\gamma_{ij}
\nonumber\\
&\quad\quad+2W^2\epsilon^k{}_{v(i}B_{j)k}
\,,\nonumber\\
\BN_{ij}&=W^2B_{ij}-W^2\epsilon^l{}_{ij}E_{lv}
-W^2\epsilon^l{}_{vi}E_{jl}\,,\label{eqn:Elec_Mag_Up_v_Low}
\end{align}
where~$\epsilon_{bcd}=n^a\epsilon_{abcd}$ stands for the lower case 
spatial volume form. Furthermore this equation shows trivially 
that changes of coordinates can not create nor destroy gravitational 
waves. Since in vacuum the electric and magnetic parts also satisfy 
a closed evolution system~\cite{NicOweZha11}, it is also clear that if 
we are given initial data with vanishing~$E_{ij}$ and~$B_{ij}$ this will 
be the case once and for all. The 
relationship~\eqref{eqn:Elec_Mag_Up_v_Low} holds in general, but using 
the vacuum Einstein equations, we have that, 
\begin{align} 
E_{ab}&=R_{ab}^{\textrm{TF}}-(K^c{}_aK_{bc})^{\textrm{TF}}+KK_{ab}^{\textrm{TF}}\,,
\end{align}
and likewise for the upper case Electric part. We can thus relate the 
tracefree part of the upper case and lower case spatial Ricci tensors
as we did for the Ricci scalars, namely by expanding out the projected 
upper case extrinsic curvature with~\eqref{eqn:KNn} and 
using~\eqref{eqn:full_from_proj} to obtain the non-spatial 
components. 

\paragraph*{Boost metric connection and curvature:} We 
would like to have the equations of motion for upper case 
objects in the lower case coordinates. But as it is more convenient 
to work with spatial tensors in the lower case coordinates we 
instead work with the boost metric and boost extrinsic 
curvature~$(\mathbbmss{g}_{ij},\mathbbmss{K}_{ij})$ to obtain the desired 
results. The time derivative of the boost vector is conveniently 
encoded as,
\begin{align}
\p_t(Wv_i)&=\alpha\,W\big[\mathbbmss{A}_i-\mathbbmss{A}_vv_i
-\mathbbmss{D}_i\ln(\alpha\,W)\big]+\Lie_\beta(Wv_i)\,.
\label{eqn:BoostvDot}
\end{align}
The relationship between the connection of the boost metric and 
the lower case spatial curvature is~$C^k{}_{ij}=
\mathbbmss{G}^k{}_{ij}-\Gamma^k{}_{ij}$, with
\begin{align}
C^k{}_{ij}&=W^2v_{(i}D_{j)}v^k-\tfrac{1}{2}D^k(W^2v_iv_j)
+\tfrac{1}{2}v^k\Lie_v\mathbbmss{g}_{ij}\nonumber\\
&=(\mathbbmss{g}^{-1})^{kl}\Big[v_{(i}\mathbbmss{D}_{j)}(W^2v_l)
-\tfrac{1}{2}\mathbbmss{D}_l(W^2v_iv_j)\Big]
\nonumber\\
&\quad+\tfrac{1}{2}v^k\Lie_v\gamma_{ij}\,.
\end{align}
This result can be obtained from the standard expression, see 
Ch.~$7$ of~\cite{Wal84}, since the lower case spatial metric and boost 
metric act in the same tangent space; notice that this is not the case 
when we try to relate the upper and lower case spatial connections. We 
could now examine how spatial geodesics are deformed in the boost metric, 
but since these are not often used in practice we elect not to do so 
here. The standard expression,
\begin{align}
\mathbbmss{R}_{ij}&=R_{ij}-2D_{[i}C^k{}_{k]j}+2C^l{}_{j[i}C^k{}_{k]l}
\nonumber\\
&=R_{ij}-2\mathbbmss{D}_{[i}C^k{}_{k]j}-2C^l{}_{j[i}C^k{}_{k]l}\,,\label{eqn:BoostRvsR}
\end{align}
similarly relates the two curvatures. Note that in the three spatial 
dimensions of the spatial slice the Weyl tensor is identically zero, so 
we need only consider the Ricci tensors rather than the full spatial 
Riemann tensors. Projecting upper case covariant derivatives into 
the lower case foliation immediately reveals the geometric meaning 
of the boost covariant derivative. Let~$S_a$ and~$S_{ab}$ denote 
upper case spatial tensors, related in the standard 
way~\eqref{eqn:full_from_proj} to their projections~$s_a$ and~$s_{ab}$ 
into the lower case foliation. Then we have,
\begin{align}
\perp\!\DN_iS&=D_iS+Wv_i\,(\Lie_NS)\,,
\end{align}
for a scalar~$S$, and,
\begin{align}
\perp\!\DN_iS_j&=\mathbbmss{D}_is_j+Wv_i\,(\Lie_Ns_j)-X^k{}_{ij}s_k\,,\nonumber\\
\perp\!\DN_iS_{jk}&=\mathbbmss{D}_is_{jk}+Wv_i\,(\Lie_Ns_{jk})
-X^l{}_{ij}s_{lk}-X^l{}_{ik}s_{jl}\,,\label{eqn:Proj_Upper_CovD}
\end{align}
for the tensors, with, 
\begin{align}
X^k{}_{ij}&=W(\mathbbmss{g}^{-1})^{kl}
\Big[(\perp\!\!\KN)_{ij}v_l-2\,(\perp\!\!\KN)_{l(i}v_{j)}\Big]\nonumber\\
&=W^2(\mathbbmss{g}^{-1})^{kl}
\big[\mathbbmss{K}_{ij}v_l-2\,\mathbbmss{K}_{l(i}v_{j)}
+v_iv_j\mathbbmss{A}_l\big]\,.
\label{eqn:boost_upper_diff}
\end{align}
The general pattern can be read off from these equations. The boost
covariant derivative is the part of the projected upper case derivative
formed from lower case spatial derivatives and the boost vector. The
remainder depends on the Lie derivative along~$N^a$ and the boost 
extrinsic curvature~$\mathbbmss{K}_{ij}$. We can relate the projected 
upper case spatial Ricci tensor and the boost curvature by 
straightforward, albeit tedious, direct computation,
\begin{align}
\mathbbmss{R}_{ij}&=\,\perp\!\!{}^{\textrm{\tiny{(N)}}}\!R_{ij}
+\mathbbmss{D}_kX^k{}_{ij}
-\mathbbmss{D}_iX^k{}_{jk}
-X^k{}_{kl}X^l{}_{ij}\nonumber\\
&\quad+X^k{}_{il}X^l{}_{jk}-Wv_i\perp(\DN_j\KN+A_j\KN)
\nonumber\\
&\quad-W^{-1}v^k\perp\big[\DN_k\KN_{ij}-2\DN_{(i}\KN_{j)k}\big]\nonumber\\
&\quad-W^{-1}v^k\perp\big[A_k\KN_{ij}-2A_{(i}\KN_{j)k}\big]\,.
\label{eqn:BoostRvsRN}
\end{align}
This result holds regardless of the gravitational field 
equations, but possible further manipulation is possible by the 
addition of the hypersurface constraints. The projected upper case 
covariant derivatives here can be replaced 
using~\eqref{eqn:Proj_Upper_CovD}, which results in a slightly 
longer expression in terms of~$\mathbbmss{K}_{ij}$.

\paragraph*{Dual foliation formulation of the wave-equation:}
As a simple example we consider a~$3+1$ decomposition of the 
wave equation~$\Box\phi=0$ using the dual foliation. For the 
Lie-derivative of the boost metric we define,
\begin{align}
\mathbbmss{P}_{ij}&\equiv\tfrac{1}{2}W\Lie_{(W^{-1}v)}\mathbbmss{g}_{ij}\,,
\end{align}
again with the convention 
that~$\mathbbmss{P}\equiv(\mathbbmss{g}^{-1})^{ij}\mathbbmss{P}_{ij}$. 
Introducing the reduction variable~$\Lie_N\phi=W\pi$. We then obtain,
\begin{align}
\p_t&\phi=\alpha\,\pi+\Lie_{(\beta-\alpha v)}\phi\,,\nonumber\\
\p_t&\pi=\alpha\,(\mathbbmss{g}^{-1})^{ij}\,
\big[\mathbbmss{D}_i\mathbbmss{D}_j\phi-X^k{}_{ij}\mathbbmss{D}_k\phi
+\mathbbmss{A}_i\,\mathbbmss{D}_j\phi\big]\nonumber\\
&+\alpha\big[\mathbbmss{K}+\mathbbmss{K}_{vv}+\mathbbmss{P}
+\Lie_{v}\log(W\alpha)\big]\,\pi 
+\Lie_{(\beta+\alpha v)}\pi\,.
\end{align}
These equations serve as a prototype when looking at the more 
complicated systems that follow. In particular the Lie derivative
terms for~$\pi$ differs from what one might naively expect.

\paragraph*{Gravitational field equations:} We 
denote~$\mathbbmss{A}_{(i}v_{j)}\equiv \mathbbmss{A}\!\otimes\!v_{ij}$.
Moving now to write the field equations in terms of the boost 
metric we obtain,
\begin{align}
\p_t\mathbbmss{g}_{ij}&=-2\,\alpha\,\mathbbmss{K}_{ij}
         + 2\,\alpha\,\mathbbmss{A}_{(i}v_{j)}  
         + \Lie_{(\beta-\alpha v)}\mathbbmss{g}_{ij}\,,\label{eqn:gNnDot}
\end{align}
and after delicate use of the first hypersurface
constraint~\eqref{eqn:Hyp_Cons},
\begin{align}
&\p_t\,\mathbbmss{K}_{ij}=\alpha\,\mathbbmss{R}_{ij}
-W^{-1}\mathbbmss{D}_{(i}\big[W^{-1}\alpha\,\mathbbmss{A}_{j)}\big]
+v^2\alpha\,\mathbbmss{A}_i\mathbbmss{A}_j\nonumber\\
&\quad\,-W^2\alpha\,(\mathbbmss{g}^{-1})^{kl}
\big(\mathbbmss{D}_k\mathbbmss{A}_l+(2-v^2)\,
\mathbbmss{A}_k\mathbbmss{A}_l\big)v_iv_j\nonumber\\
&\quad\,-\alpha\,\Lie_{(W^{-1}v)}\big(W\mathbbmss{A}\!\otimes\!v_{ij}\big)
-2\,W^{-2}\mathbbmss{D}_{(i}\big[W^2\alpha\,\mathbbmss{K}_{j)v}\big]\nonumber\\
&\quad\,
-2\,W^2\alpha\,(\mathbbmss{g}^{-1})^{kl}(\mathbbmss{g}^{-1})^{m}{}_{(i}v_{j)}
\mathbbmss{A}_k\mathbbmss{K}_{lm}-W^2\alpha\,\mathbbmss{A}_v\mathbbmss{K}v_iv_j
\nonumber\\
&\quad\,-2\,\alpha\,\mathbbmss{A}_{(i}\mathbbmss{K}_{j)v}
+\alpha\,\big(\mathbbmss{K}+\mathbbmss{K}_{vv}+\mathbbmss{P}
+\Lie_{v}\log(W\alpha)\big)\,\mathbbmss{K}_{ij}
\nonumber\\
&\quad\,+\alpha\,\mathbbmss{K}\,\mathbbmss{P}_{ij}
-\alpha\,\big(v^2\mathbbmss{K}+2\,\mathbbmss{K}_{vv}+\mathbbmss{P}
+W^{-2}\mathbbmss{A}_v\big)\,\mathbbmss{A}\!\otimes\!v_{ij}\nonumber\\
&\quad\,-2\,\alpha\,(\mathbbmss{K}-\mathbbmss{A}\!\otimes\!v)^k{}_{i}\,
(\mathbbmss{K}-\mathbbmss{A}\!\otimes\!v)_{jk}
-2\,\mathbbmss{K}_{v(i}\,\mathbbmss{D}_{j)}\alpha\nonumber\\
&\quad\,-4\,\alpha\,(\mathbbmss{g}^{-1})^{kl}(\mathbbmss{K}-\mathbbmss{A}\!\otimes\!v)_{k(i}
\mathbbmss{P}_{j)l}+\Lie_{(\beta+\alpha v)}\mathbbmss{K}_{ij}\,.\label{eqn:KNnBDot}
\end{align}
Notice that we end up here with equations involving principal derivatives 
of the lower case shift but not the lower case lapse. That is, in equation~\eqref{eqn:gNnDot}
first derivatives of~$\beta^i$ appear, but in equation~\eqref{eqn:KNnBDot} no second 
derivatives of~$\alpha$ appear, and instead we have derivatives of~$\mathbbmss{A}_i$. This 
should be compared with the form of the equation~\eqref{eqn:ADM_eqns} in which second derivatives
of~$\alpha$ are present. Intuitively this happens because we are mixing the use of lower case 
spatial coordinates with upper case spatial objects. The Hamiltonian constraint can 
also be expressed in terms of these variables, giving,
\begin{align}
\mathbbmss{H}&=
\mathbbmss{R}
+(\mathbbmss{K}+\mathbbmss{K}_{vv})^2-\mathbbmss{K}_{ij}\,\mathbbmss{K}^{ij}
+2\,\mathbbmss{D}_{i}\big(v^i\,\mathbbmss{K}\big)\nonumber\\
&\quad-2\,(\mathbbmss{g}^{-1})^{ij}\mathbbmss{D}_{i}\big(\mathbbmss{K}_{jv}+W^{-1}v^k\,\mathbbmss{D}_{[k}Wv_{j]}\big)
\nonumber\\
&\quad+W^{-2}\,\gamma^{ij}\gamma^{kl}\,\mathbbmss{D}_{[i}Wv_{k]}\mathbbmss{D}_{[j}Wv_{l]}\,.\label{eqn:HamnB}
\end{align}
Likewise the momentum constraint becomes,
\begin{align}
&\mathbbmss{M}_i=
W(\mathbbmss{g}^{-1})^{jk}\,\mathbbmss{D}_j(W^{-1}\mathbbmss{K}_{ki})
-\mathbbmss{D}_i(\mathbbmss{K}+\mathbbmss{K}_{vv})
+\Lie_v\mathbbmss{K}_{vi}
\nonumber\\
&\,\,+W(\mathbbmss{g}^{-1})^{jk}\big[\mathbbmss{D}_j\mathbbmss{D}_{[k}Wv_{i]}
-\tfrac{1}{2}Wv_i\,\mathbbmss{D}_{j}
(W^{-1}v^l\,\mathbbmss{D}_{[l}Wv_{k]})\big]\nonumber\\
&\,\,-2(\mathbbmss{g}^{-1})^{jk}v^l\,\mathbbmss{D}_{[l}Wv_{j]}\mathbbmss{D}_{[k}Wv_{i]}
+\mathbbmss{R}_{iv}
+\mathbbmss{P}\,\mathbbmss{K}_{vi}\,,\label{eqn:MomnB}
\end{align}
where, up to additions of the hypersurface constraints, we 
define~$\mathbbmss{H}=H-2M_v$ and~$\mathbbmss{M}_i=M_i$. For readability
in equations~\eqref{eqn:HamnB} and~\eqref{eqn:MomnB} we 
write~$\mathbbmss{D}_{[i}Wv_{j]}\equiv\mathbbmss{D}_{[i}(Wv_{j]})$. No 
complications arise in including the stress-energy tensor for nonvacuum
spacetimes. We note in passing that by taking~$\mathbbmss{g}_{ij}$ 
and~$\mathbbmss{K}_{ij}$ as evolved variables the evolution equation for 
the metric also looks natural, since the projected upper case acceleration 
appears as it would in a tetrad formulation when the timelike leg differs 
from the normal vector~$n^a$. With this choice of variables, we also see 
that the constraints can be written so that there is no explicit appearance 
of the projected upper case acceleration~$\mathbbmss{A}_i$. Furthermore, 
there is no explicit appearance of~$\Lie_N\mathbbmss{A}_i$ in 
equation~\eqref{eqn:KNnBDot}. Nevertheless we may be more interested in 
how the projected upper case extrinsic curvature evolves. This slightly 
more compact expression is trivially obtained by 
combining~\eqref{eqn:KNnBDot} and~\eqref{eqn:BoostvDot}. 
The equations are given in both forms in mathematica notebooks that accompany 
the paper~\cite{HilWebsite}. One expects to be able to formulate an initial 
data construction strategy naturally around the boost metric. A natural starting 
point for this would be to examine conformal flatness of the boost metric in 
boosted Schwarzschild. See~\cite{RucHeaLou14} for work along these lines. 
Observe that the notation in these last equations is made slightly more cumbersome 
by continuing to insist on raising and lowering indices with the spatial 
metric~$\gamma_{ij}$, but we prefer to do so to avoid confusion with the 
surrounding calculations. 

\paragraph*{The Generalized Jang equation:} The Jang 
equation~\cite{Jan78} is a quasilinear partial differential 
equation of minimal surface type, originally introduced as a tool
for the proof of the positive energy theorem. Given initial 
data~$(\gamma_{ij},K_{ij})$ for the initial value problem in 
General Relativity it reads,
\begin{align}
\left(\gamma^{ij}-\frac{D^iTD^jT}{1+D^kTD_kT}\right)
\left(K_{ij}-\frac{D_iD_jT}{(1+D^kTD_kT)^{1/2}}\right) =0\,,
\label{eqn:Jang}
\end{align}
for the unknown scalar field~$T$. This equation was motivated 
by the characterization of slices of the Minkowski spacetime,
in which there exists a scalar function~$T$ such that the second
bracket of~\eqref{eqn:Jang} vanishes, and such that the boost 
metric,
\begin{align}
\mathbbmss{g}_{ij}=\gamma_{ij}+D_iT\,D_jT\,,\nonumber 
\end{align}
is flat. With the present formalism it is clear that the natural 
curved space generalization to~\eqref{eqn:Jang} should be,
\begin{align}
(\mathbbmss{g}^{-1})^{ij}\,(\mathbbmss{K}_{ij}-\mathbbmss{A}_{(i}v_{j)})
=\,\mathbbmss{K}-W^{-2}\mathbbmss{A}_{v}=0\,,
\end{align}
which, remarkably corresponds to the upper case 
foliation being maximal, because using the inverse boost metric to trace 
projected upper case quantities reveals the full trace of the original 
upper case tensor. One furthermore expects an analogous characterization 
of general asymptotically flat initial data sets to that of flat-space, in 
roughly the following terms: Consider data~$(\gamN_{\ul{ij}},\KN_{\ul{ij}})$ 
extracted from a spacetime~$(M,g)$, written in 
coordinates~$X^{\ul{\mu}}=(T,X^{\ul{i}})$, on some Cauchy slice, not necessarily 
a level set of~$T$. An initial data set~$(\gamma_{ij},K_{ij})$ corresponds to 
the same data if and only if there exist vectors~$(v^i,\,\mathbbmss{A}_i)$ 
satisfying the hypersurface constraint~\eqref{eqn:Hyp_Cons}, which we may 
write in the form,
\begin{align}
(D\times Wv)_i&=(D\ln A\,\times A^{-1}Wv)_i\,,
\end{align}
and a {\it projected Jacobian transformation~$\varphi^{\ul{i}}{}_i$} with 
inverse~$(\varphi^{-1})^{i}{}_{\ul{i}}$ such that,
\begin{align}
\mathbbmss{g}_{ij}&=\gamma_{ij}+W^2v_iv_j\,,\nonumber\\
\mathbbmss{K}_{ij}&=K_{ij}-W^{-1}D_{(i}Wv_{j)}\,,
\end{align}
satisfy,
\begin{align}
\gamN_{\ul{ij}}&=(\varphi^{-1})^i{}_{\ul{i}}\,(\varphi^{-1})^j{}_{\ul{j}}
\,\mathbbmss{g}_{ij}\,,
\nonumber\\
\KN_{\ul{ij}}&=W(\varphi^{-1})^i{}_{\ul{i}}\,(\varphi^{-1})^j{}_{\ul{j}}\,
(\mathbbmss{K}_{ij}-\mathbbmss{A}_{(i}v_{j)})\,,
\end{align}
everywhere on the constant-$t$ slice. The relationship between the projected 
and true Jacobian transformation is that,
\begin{align}
\varphi^{\ul{i}}{}_i&=\perpN^{\ul{i}}\!\!_{\ul{\mu}}J^{\ul{\mu}}{}_i\,, &\quad
(\varphi^{-1})^i{}_{\ul{i}}=\perp^i\!\!_\mu (J^{-1})^{\mu}{}_{\ul{i}}+Wv^iV_{\ul{i}}\,.
\label{eqn:J_proj}
\end{align}
The transformation~$\varphi^{\ul{i}}{}_i$ has the property that it 
maps N-spacelike contravariant tensor indices in~$X^{\ul{i}}$ coordinates 
to~$x^i$ coordinates and simultaneously projects the object into the 
lower case slice, and vice-versa for covariant n-spacelike indices in 
coordinates~$x^i$. The inverse property is easily verified by direct 
computation. The equivalent transformation~$\varPhi^i{}_{\ul{i}}$ in 
the opposite direction is defined in the obvious way. In the special case 
that the upper case coordinates are global inertial on Minkowski spacetime 
this characterization reduces to that stated above motivating the Jang 
equation. We propose that this, rather than the conformal transformation, 
as is sometimes claimed, constitutes the relation that should be used 
to build the natural equivalence class over physically equivalent 
solutions to metric-based formulations of GR.

\paragraph*{Discussion of the dual foliation initial value 
problem:} Given a boost metric~$\mathbbmss{g}_{ij}$, projected 
extrinsic curvature~$\mathbbmss{K}_{ij}$, boost vector~$v_i$, and 
acceleration~$\mathbbmss{A}_i$ satisfying the hypersurface, Hamiltonian 
and momentum constraints we have a suitable set of initial data for 
vacuum GR. None of these quantities are invariant under changes of 
the upper case time coordinate~$T$, but the form of the field equations 
is nevertheless invariant under this change. One way to view the 
resulting additional freedom is that by breaking the correspondence 
between the spatial metric and the projection operator onto slices 
of the foliation, we gain the freedom to take the spatial metric of 
other foliations as the evolved variable. The subsequent projection 
into the foliation to obtain the boost metric is the most convenient 
way to deal with the variable in the~$3+1$ language, and fits nicely 
with earlier work such as the Jang equation. It is natural to compare 
this reformulation with the freedom in the Maxwell equations, whose 
gauge can be altered without changing the coordinates on spacetime. 
The electric and magnetic fields are invariant under such changes, as 
they depend only on the Faraday tensor and the choice of coordinates. We 
have exactly the same status with the boost freedom; the electric and 
magnetic parts of the Weyl tensor are determined purely by the choice 
of coordinates, according to~\eqref{eqn:E_B_defn}, and thus independent 
of the  choice in the boost freedom. But the form of the field equations 
is invariant under changes to the boost. 

\paragraph*{Working with the upper case spatial tensor basis:} Allowing 
the boost freedom decouples, in the highest derivatives, the lower 
case lapse from the evolved variables. Therefore it is natural to ask 
whether such a decoupling can also be obtained in the lower case 
shift by keeping all tensors in the~$X^{\ul{i}}$ coordinate basis.
We therefore now wish to drop the Jang-equation style use of two 
time coordinates with everything expressed in the coordinate 
basis~$x^i$ vectors, and instead use the~$X^{\ul{i}}$ basis tensor 
components, whilst computing derivatives in the~$x^\mu$ coordinates.
The time derivative of the projected Jacobian is,
\begin{align}
\p_t\varphi^{\ul{i}}{}_i&=
\Lie_{(\beta-\alpha v)}\varphi^{\ul{i}}{}_i
-\alpha\,(AW)^{-1}\varphi^{\ul{j}}{}_i\p_{{\ul{j}}} B^{\ul{i}}
\,,\nonumber\\
\p_t(\varphi^{-1})^{i}{}_{\ul{i}}&=\Lie_{(\beta-\alpha v)}(\varphi^{-1})^{i}{}_{\ul{i}}
+\alpha\,(AW)^{-1}(\varphi^{-1})^{i}{}_{\ul{j}}
\p_{{\ul{i}}} B^{\ul{j}}\,.\label{eqn:Proj_Jac_Dot}
\end{align}
Given the time derivative of an upper case spatial tensor in lower 
case coordinates we can now use~\eqref{eqn:Proj_Jac_Dot} to compute 
the lower case time derivative in the upper case basis. Take for 
example~$S_{ab}$ symmetric upper case spatial, again with projection~$s_{ij}$
into the lower case foliation. Suppose we have,
\begin{align}
\p_ts_{ij}&=\alpha\,X_{ij}+\Lie_\beta s_{ij}\,,
\end{align}
then it follows that,
\begin{align}
\p_tS_{\ul{ij}}&=\alpha\,(\varphi^{-1})^i{}_{\ul{i}}(\varphi^{-1})^j{}_{\ul{j}}X_{ij}
+2\,\alpha\,(AW)^{-1} S_{\ul{k}(\ul{i}}\p_{\ul{j})}B^{\ul{k}}\nonumber\\
&\quad+\Lie_{(\beta-\alpha v)}S_{\ul{ij}}\,.
\end{align}
Taking~$S_{ab}$ as the upper case spatial metric, and looking at~\eqref{eqn:gNnDot} 
we immediately see that indeed the lower case shift does become decoupled in 
the highest derivatives. It is sufficient to consider only this 
evolution equation because this is the only place where the shift 
is coupled in the principal part. The relationship between the upper case 
Christoffel symbol and that of the boost metric is,
\begin{align}
&\varphi^{\ul{i}}{}_i\,\varphi^{\ul{j}}{}_j\,
(\varphi^{-1})^{k}{}_{\ul{k}}
\,{}^{\textrm{\tiny{(N)}}}\Gamma^{\ul{k}}{}_{\ul{ij}} =
\mathbbmss{G}^k{}_{ij}+X^k{}_{ij}\nonumber\\
&\quad\quad\quad\,\,+(\varphi^{-1})^k{}_{\ul{k}}\big[
\p_{(i}\varphi^{\ul{k}}{}_{j)}-A^{-1}Wv_{(i}\varphi^{\ul{l}}{}_{j)}\p_{\ul{l}}B^{\ul{k}}\big]
\,,\label{eqn:conn_boost_upper}
\end{align}
with~$X^k{}_{ij}$ defined as above~\eqref{eqn:boost_upper_diff}.
Note that~\eqref{eqn:conn_boost_upper} can be rewritten as,
\begin{align}
\mathbbmss{D}_{(i}\varphi^{\ul{k}}{}_{j)}
+\!{}^{\textrm{\tiny{(N)}}}\Gamma^{\ul{k}}{}_{\ul{ij}}
\,\varphi^{\ul{i}}{}_i\,\varphi^{\ul{j}}{}_j=X^k{}_{ij}\,\varphi^{\ul{k}}{}_k
-A^{-1}Wv_{(i}\varphi^{\ul{l}}{}_{j)}\p_{\ul{l}}B^{\ul{k}}\,,
\end{align}
which in the setting with vanishing boost vector can be interpreted as 
the statement that the Levi-Civita connection of the spatial metric is 
the gauge covariant derivative associated with spatial diffeomorphisms. 
The full field equations are given in this mixed basis form 
in~\cite{Hil15}, together with a canonical Hamiltonian treatment.

\section{Double-null formulation}
\label{section:DN}

In this section we work in coordinates~$x^{\mu}=(t,x^i)$. Throughout 
latin indices~$a,b,c,d,e$ will be abstract as in the previous
section, and likewise latin indices~$i,j,k,l,m,p$ stand for 
spatial components in~$x^\mu$ as before. We perform a~$2+1$ 
decomposition against~$r$ on the spatial slice, and 
take~$x^\mu=(t,r,\theta^A)$ to be adapted coordinates. Thus upper 
case latin indices~$A,B,C,D$ stand for those in the level-sets 
of~$r$.

\subsection{$2+1+1$ Decomposition}\label{subsection:2+1+1}

\paragraph*{Motivation:} We now turn our attention towards 
finding coordinates suitable for studying the collapse of 
gravitational waves. It is known that sufficiently small 
perturbations of the Minkowski spacetime are long-lived in the 
pure harmonic gauge~\cite{LinRod04,LinRod05}, but this class of 
initial data presumably does not include every possible data set 
that eventually asymptote to the Minkowski spacetime; for 
sufficiently strong data, or indeed sufficiently strong pure gauge 
perturbations, coordinate singularities are expected to form. 
Indeed there are examples of this phenomenon~\cite{Her00}, and 
it may be that some of the difficulties in evolving strong Brill 
waves in~\cite{Rin06} were caused by the use of the pure harmonic 
slicing. Therefore we look elsewhere. Empirically the generalized 
harmonic gauges~\cite{SziLinSch09} have been found very robust in both 
binary blackhole and collapse scenarios. However, from the 
mathematical point of view, the strongest results concerning collapse 
to a blackhole employ a double-null foliation~\cite{Chr08}. It is 
known that a particular type of initial data will form an apparent 
horizon before any coordinate singularity forms. It is not clear 
that close to the critical threshold of blackhole formation these 
coordinates are well-behaved. It is also rather doubtful that the 
double-null foliation will be useful in the strong-field region in 
binary-blackhole spacetimes. But since these coordinates naturally 
conform to the causal structure of the spacetime, and there is 
likewise no guarantee of nice behavior for any other coordinate 
system, they seem to be in the best shape for consideration. In 
particular, in both~$3+1$-spherical symmetry~\cite{Gar95} and a~$2+1$ 
dimensional setting~\cite{PreCho00} such coordinates have been effectively 
used in studies of critical collapse, neatly sidestepping the need 
for mesh-refinement. We thus look at related gauges suitable for 
the initial value problem.

\paragraph*{The~$2+1$ decomposition:} In a double-null foliation
there are two crucial coordinates, optical functions, whose 
level sets are incoming and outgoing null surfaces. In the~$3+1$
setting we have however only singled out the time-coordinate
for special treatment. A given pair of these null hypersurfaces 
intersect in a spacelike two-sphere. Given a spatial 
slice of constant~$t$, complete with spatial metric~$\gamma_{ij}$ and 
extrinsic curvature~$K_{ij}$ let us define a new coordinate~$r$, 
which we will use to perform a~$2+1$ split. The idea is that 
the level sets of~$r$ should become the spheres on which the 
null surfaces intersect. Obviously the calculations that follow 
are essentially the same as those of the standard~$3+1$ split, 
as described in detail in~\cite{Gou07}. The coordinate~$r$ defines 
a unit normal~$s^i$ to a surface of constant~$r$ according to,
\begin{align}
L^{-2}&=\gamma^{ij}(D_ir)(D_jr)\,,\quad 
s^i=\gamma^{ij}LD_jr\,.
\end{align}
We will call~$L$ the length scalar. The normal vector~$s^i$ 
naturally defines the induced metric in the two-dimensional 
level set,
\begin{align}
q_{ij}&=\gamma_{ij}-s_is_j\,.
\end{align}
Likewise we have the extrinsic curvature,
\begin{align}
\chi_{ij}&=q^k{}_iD_ks_j\,,
\end{align}
so the first derivatives of the spatial normal vector are in 
total,
\begin{align}
D_is_j&=\chi_{ij}-s_i\sD_j\ln L\,.
\end{align}
We denote the covariant derivative compatible with the induced 
metric~$q_{ij}$ by~$\sD_i$, and likewise use~$\sP_i$ for the partial 
derivative projected into the surface. Note the relative change in 
sign in the definition of this extrinsic curvature and that of the 
slice~$K_{ij}$. Let the vector~$r^i$ be tangent to lines of 
constant~$\theta^A$, the two spatial coordinates in the level set. 
We have 
\begin{align}
r^i&=L s^i + b^i\,,
\end{align}
with~$b^is_i=0$. We call the two-dimensional vector~$b^i$ the 
slip vector. Note the relation~$r^iD_ir=1$. With this notation 
we can express the spatial metric as,
\begin{align}
\textrm{d}l^2=L^2\textrm{d}r^2+q_{AB}
(\textrm{d}\theta^A+b^A\textrm{d}r)
(\textrm{d}\theta^B+b^B\textrm{d}r)\,.
\end{align}
When performing the~$3+1$ decomposition one finds that the 
lapse scalar and shift vector are freely specifiable, which 
is of course not the case with the analogous length scalar and 
slip vector. The four-dimensional metric is,
\begin{align}
\textrm{d}s^2&=-\alpha^2\textrm{d}t^2+L^2(\textrm{d}r+L^{-1}\beta^s\textrm{d}t)^2
\nonumber\\
&+q_{AB}
(\textrm{d}\theta^A+b^A\textrm{d}r+\beta^A\textrm{d}t)
(\textrm{d}\theta^B+b^B\textrm{d}r+\beta^B\textrm{d}t)\,.
\end{align}
The coordinate light speeds in the increasing and decreasing~$r$ 
directions are,
\begin{align}
c^r_\pm&=(-\beta^s\pm\,\alpha)\,L^{-1}\,,\label{eqn:coord_light}
\end{align}
whilst in the transverse directions we have, 
\begin{align}
c^A_\pm&=-\beta^A\mp\,b^A\,\alpha\,L^{-1}\,.\label{eqn:coord_light_trans}
\end{align}
Although they may not necessarily be associated with spherical-polar 
coordinates we will call these transverse directions `angular'. An obvious 
choice is~$\alpha=L$, and~$\beta^s=0$, under which the coordinate 
light-speeds are~$c^r_\pm=\pm1$. With this choice the combinations~$u=t-r$ 
and~$v=t+r$ are the optical outgoing and incoming null coordinates alluded 
to earlier, and the coordinates are naturally adapted to the causal 
structure of the spacetime in the null~$n^a\pm s^a$ directions. 

\paragraph*{The extrinsic curvatures:} We immediately split the 
extrinsic curvature~$\chi_{AB}$ into a trace and tracefree part,
\begin{align}
\chi_{AB}&=\hat{\chi}_{AB}+\tfrac{1}{2}q_{AB}\,\chi\,.
\end{align}
We will similarly use the notation~$q$ for the determinant of 
the two-metric. Finally we decompose the extrinsic curvature of 
the spacelike surface as embedded in the spacetime by 
\begin{align}
K_{ij}&=s_is_jK_{ss}+\tfrac{1}{2}q_{ij}K_{qq}+2s_{(i}q^A{}_{j)}K_A
+\hat{K}_{AB}\,,\label{eqn:K_2+1}
\end{align}
where in accordance with the previous notation~$\hat{K}_{AB}$ 
stands for the projected, tracefree part. We use indices~$s$
to denote contraction with~$s^a$, and~$qq$ for a trace taken with 
the two-dimensional metric~$q_{AB}$. We 
write~$K_{AB}=q^i{}_Aq^j{}_BK_{ij}$ for the projected 
extrinsic curvature, and occasionally still use~$K=K_{ss}+K_{qq}$ 
as a shorthand for the trace of the extrinsic curvature.

\paragraph*{The Christoffel symbol:} The spatial Christoffel 
symbol is readily decomposed as,
\begin{align}
\Gamma^s{}_{ss}&=\Lie_s(\ln L)\,,\quad
\Gamma^s{}_{si}\,q^i{}_A=\sP_A(\ln L)\,,\nonumber\\
\Gamma^s{}_{ij}\,q^i{}_A\,q^j{}_B&=-\chi_{AB}\,,\quad
\Gamma^k{}_{ij}\,q^i{}_A\,q^j{}_B\,q_{k}{}^C=\sG^C{}_{AB}\,,\nonumber\\
\Gamma^k{}_{ss}\,q^A{}_k&=-\sP^A(\ln L)
+\tfrac{1}{L^2}\left(\Lie_rb^A-b^B\sP_Bb^A\right)\,,
\nonumber\\
\Gamma^k{}_{is}\,q^i{}_B\,q_{kA}\,&=\chi_{AB}+\tfrac{1}{L}q_{AC}\sP_{B}b^C\,,
\end{align}
with~$\sG$ denoting the Christoffel symbols of the two-metric~$q_{AB}$.
The spatial contracted Christoffel symbols~$\Gamma^i=\gamma^{jk}\Gamma^i{}_{jk}$ 
are therefore,
\begin{align}
\Gamma^s&=\Lie_s(\ln L)-\chi\,,\nonumber\\
\Gamma^iq_i{}^A&=\sG^{\,A}+\tfrac{1}{L^2}\left(\Lie_rb^A
-b^B\sP_Bb^A\right)-\sD^A(\ln L)\,.
\end{align}

\paragraph*{Curvature:} As in~\eqref{eqn:K_2+1} the spatial 
Ricci curvature~$R_{ij}$ can be decomposed according to,
\begin{align}
R_{ss}&=-\tfrac{1}{L}\sD^{\,A}\sD_AL-\Lie_s\chi-\chi^{A}{}_{B}\chi^B{}_{A}
\,,\nonumber\\
R_{qq}&=-\tfrac{1}{L}\sD^{\,A}\sD_AL-\Lie_s\chi+\sR-\chi^2\,,
\nonumber\\
R_{sA}&=\sD_B\chi^B{}_A-\sD_A\chi\,,\nonumber\\
\hat{R}_{AB}&=-\Lie_s\hat{\chi}_{AB}-\tfrac{1}{L}\sD_A\sD_B^{\,TF}L
+2\,\hat{\chi}^C{}_A\,\hat{\chi}_{BC}\,,
\end{align}
by the classical Gauss-Codazzi-Mainardi equations.

\paragraph*{Vacuum field equations:} The $2+1+1$ decomposed 
Einstein equations are given first by the constraints,
\begin{align}
H&=\sR-\tfrac{2}{L}\sD_A\sD^{\,A}L-2\Lie_s\chi-\hat{\chi}_{AB}\hat{\chi}^{AB}-\tfrac{3}{2}
\chi^2\nonumber\\
&\quad-\tfrac{1}{2}K_{qq}^2-2K_{qq}K_{ss}+2K_AK^A+\hat{K}_{AB}\hat{K}^{AB}\,,\nonumber\\
M_s&=-\Lie_sK_{qq}+\sD_AK^A+2K^A\sD_A(\ln L)+\chi K_{ss}\nonumber\\
&\quad-\tfrac{1}{2}\chi K_{qq}+\hat{\chi}^A{}_B \hat{K}^B{}_A\,,\nonumber\\
M_A&=\Lie_sK_A+\sD^{\,B}\hat{K}_{AB}-\tfrac{1}{2}\sD_AK_{qq}-\sD_AK_{ss}+\chi K_A
\nonumber\\
&\quad-K_{ss}\,\sD_A(\ln L)+\tfrac{1}{2}K_{qq}\sD_A(\ln L)
+\hat{K}_{AB}\sD^{\,B}(\ln L)\,.\label{eqn:DN_constraints}
\end{align}
Since null geodesic expansions in the~$n^a\pm s^a$ directions are 
given by~$\chi\pm K_{qq}$, we can view the~$H$ and~$M_s$ constraints as 
dictating how the expansions vary over the slice. Next we have 
evolution equations for the metric,
\begin{align}
\p_t(\ln L)&=-\alpha K_{ss}+\beta^A\sD_A(\ln L)+D_s\beta^s\,,\nonumber\\
\p_tb^A&=-2\alpha LK^A+L^2\sD^A(L^{-1}\beta^s)+\Lie_r\beta^A+\sLie_{\beta}b^A
\,,\nonumber\\
\p_tq_{AB}&=-2\alpha K_{AB}+2\beta^s\chi_{AB}+\sLie_{\beta}q_{AB}\,,
\end{align}
and for the decomposed extrinsic curvature,
\begin{align}
\p_tK_{ss}&=-\Lie_s\Lie_s\alpha+\alpha [R_{ss}+2K_AK^A+K_{qq}K_{ss}+K_{ss}^2]
\nonumber\\
&-\sD^A(\ln L)\sD_A\alpha-2LK_{ss}D_s(L^{-1}\beta^s)\nonumber\\
&-2LK^A\sD_A(L^{-1}\beta^s)+\beta^sD_sK_{ss}+\beta^A\sD_AK_{ss}\,,\nonumber\\
\p_tK_{qq}&=
-\sD^A\sD_A\alpha+\alpha[R_{qq}+K_{qq}^2+2K_AK^A+K_{qq}K_{ss}]\nonumber\\
&-\chi\Lie_s\alpha+2LK^A\sD_A(L^{-1}\beta^s)+\beta^sD_sK_{qq}\nonumber\\
&+\beta^A\sD_AK_{qq}\,,\nonumber\\
\p_tK_A&=-\sD_A\Lie_s\alpha+\alpha[R_{sA}+K_{qq}K_A]+\chi^B{}_A\sD_B\alpha
\nonumber\\
&-LK_AD_s(L^{-1}\beta^s)-LK^B{}_A\sD_B(L^{-1}\beta^s)\nonumber\\
&+\sLie_\beta K_A+\beta^s\Lie_sK_A\,,\nonumber\\
\p_t\hat{K}_{AB}&=-\sD_A\sD_B^{\,TF}\alpha-\hat{\chi}_{AB}\Lie_s\alpha
+2LK_{(A}\sD_{B)}^{\,TF}(L^{-1}\beta^s)\nonumber\\
&+\alpha[\hat{R}_{AB}-2\hat{K}^C{}_A\hat{K}_{BC}+(K_AK_B)^{\,TF}
+K_{ss}\hat{K}_{AB}]\nonumber\\
&+\beta^s\Lie_s\hat{K}_{AB}+\sLie_\beta\hat{K}_{AB}\,.
\end{align}
Here we have defined the Lie-derivative in the level-set of~$r$, in the 
obvious way, and where it is understood that the vector argument must be
projected with~$q_{AB}$, which allows us to write, for example,
\begin{align}
\sLie_\beta\hat{K}_{AB}=q^k{}_Aq^l{}_B\big(\,q^j{}_i\beta^i\sD_j\hat{K}_{kl}
+2\hat{K}_{j(k}\sD_{j)}(q^j{}_i\beta^i\,)\,\big)\,.
\end{align}
Finally the equation of motion for the extrinsic curvature~$\chi_{ij}$
can be computed from the relation~$2\chi_{ij}=\Lie_sq_{ij}$. The 
result is,
\begin{align}
\p_t\chi_{AB}&=-\Lie_s(\alpha K_{AB})+\alpha\sLie_Kq_{AB}+2K_{(A}\sD_{B)}\alpha
\nonumber\\
&+\alpha K_{ss}\chi_{AB}+2\alpha K_{(A}\sD_{B)}(\ln L)
+(\tfrac{1}{L}\sD_A\sD_BL)\beta^s\nonumber\\
&-\sD_A\sD_B\beta^s+\beta^s\Lie_s\chi_{AB}
+\sLie_\beta\chi_{AB}\,.
\end{align}
If we want to treat~$r$ as a radial coordinate, and the 
remaining~$\theta^A$ as angular coordinates we obtain regularity 
conditions at~$r=0$. These conditions will be discussed 
elsewhere. 

\subsection{Double-null formulation}\label{subsection:DN}

We now look at the field equations imposing the double-null 
gauge explicitly. The aim here is first, to present the 
simplified form of the field equations in this gauge, and 
second, to examine whether or not hyperbolicity of the full 
system can be obtained.

\paragraph*{Pure gauge analysis:} Let us examine the behavior 
of infinitesimal perturbations to coordinates satisfying the 
optical conditions~$\alpha=L$ and~$\beta^s=0$ above, additionally 
taking~$\beta^A=b^A$. This sign is chosen assuming that   
the gravitational wave is traveling mostly in the minus~$r$
direction consistent with earlier work. This is Christodolou's 
gauge choice in~\cite{Chr08}, rewritten in a time-space rather 
than a double-null form. The expressions~\cite{HilRic13} for the 
time development of the perturbations to the time and space 
coordinates are,
\begin{align}
\p_t\theta&=U-\psi^iD_i\alpha+\beta^i\p_i\theta\,,\nonumber\\
\p_t\psi^i&=V^i+\alpha D^i\theta-\theta D^i\alpha+\Lie_\beta \psi^i\,.
\end{align}
Then we have,
\begin{align}
U&\equiv\Delta[\alpha]=\Delta[L]=\Delta[(\gamma^{ij}D_irD_jr)^{-1/2}]
\nonumber\\
&=L\,\Lie_s\psi_s-\theta LK_{ss}+\psi^A\sD_AL\,,\nonumber\\
V^s&\equiv s_i\Delta[\beta^i]\,,\nonumber\,\\
V^A&\equiv\Delta[\beta^A]=\Delta[b^A]\,\nonumber\\
&=-2\theta LK^A+L^2\sD^A(L^{-1}\psi_s)+\Lie_r\psi^A+\sLie_{\psi}b^A\,,
\end{align}
and obtain the pure gauge subsystem,
\begin{align}
\p_t[L^{-1}\theta]&=LD_s[L^{-1}\psi_s]+b^A\sD_A[L^{-1}\theta]\,,\nonumber\\
\p_t[L^{-1}\psi_s]&=LD_s[L^{-1}\theta]+b^A\sD_A[L^{-1}\psi_s]\,,\nonumber\\
\p_t(q\cdot\psi)^A&=\p_r(q\cdot\psi)^A+L^2\sD^A[L^{-1}(\theta+\psi_s)]\nonumber\\
&\quad-2LK^A(\theta+\psi_s)\,.\label{eqn:DN_PG}
\end{align}
This first order PDE system is only weakly hyperbolic. The 
arguments presented in~\cite{HilRic13}, building on those 
of~\cite{KhoNov02,HilRic10}, can be used to show that no 
strongly hyperbolic formulation can be built with this gauge condition, 
at least if the formulation is constructed under the standard 
free-evolution approach. Therefore the double-null gauge can not be 
directly used in numerical relativity in the standard way, but requires 
a more subtle approach, or some modification. Note that the problem here 
comes from the choice~$\beta^A=b^A$, and there are simple modifications 
under which strong hyperbolicity of the pure gauge subsystem can be 
obtained. Nevertheless, building a formulation of GR which is at least 
strongly hyperbolic with one of these {\it good}, modified, conditions 
will be more involved because here the~$s^i$ direction is singled out 
for special treatment. Therefore in the following sections we instead 
look for a simpler approach employing the dual foliation formalism.

\paragraph*{Fixing the gauge:} Despite the shortcomings unearthed
by the pure gauge analysis, for completeness we present the full field 
equations with in the double-null form. As above we 
choose~$\alpha=L$,~$\beta^s=0$ and~$\beta^A=b^A$. This choice has 
no effect on the constraints~\eqref{eqn:DN_constraints}, but the 
evolution equations become,
\begin{align}
\p_t(\ln L)&= - L K_{ss}+b^A\sD_A(\ln L)\,,\nonumber\\
\p_tb^A&=-2L^2K^A+\Lie_rb^A\,,\nonumber\\
\p_tq_{AB}&=-2\alpha K_{AB}+\sLie_{b}q_{AB}\,,
\end{align}
for the metric components, and
\begin{align}
\p_tK_{ss}&=-\Lie_s\Lie_sL+L [R_{ss}+2K_AK^A+K_{qq}K_{ss}+K_{ss}^2]
\nonumber\\
&\quad-\sD^A(\ln L)\sD_AL+b^A\sD_AK_{ss}\,,\nonumber\\
\p_tK_{qq}&=
-\sD^A\sD_AL+L[R_{qq}+K_{qq}^2+2K_AK^A+K_{qq}K_{ss}]\nonumber\\
&\quad-\chi\Lie_sL+b^A\sD_AK_{qq}\,,\nonumber\\
\p_tK_A&=-\sD_A\Lie_sL+L[R_{sA}+K_{qq}K_A]+\chi^B{}_A\sD_BL\nonumber\\
&\quad+\sLie_\beta K_A\,,\nonumber\\
\p_t\hat{K}_{AB}&=-\sD_A\sD_B^{\,TF}L-\hat{\chi}_{AB}\Lie_sL
+L[\hat{R}_{AB}-2\hat{K}^C{}_A\hat{K}_{BC}
\nonumber\\
&\quad+(K_AK_B)^{\,TF}+K_{ss}\hat{K}_{AB}]
+\sLie_b\hat{K}_{AB}\,,
\end{align}
for the extrinsic curvature. Finally we have,
\begin{align}
\p_t\chi_{AB}&=-\Lie_s(L K_{AB})+L\sLie_Kq_{AB}
+2K_{(A}\sD_{B)}L\nonumber\\
&\quad+L K_{ss}\chi_{AB}+2L K_{(A}\sD_{B)}(\ln L)
+\sLie_b\chi_{AB}\,.
\end{align}
Taking linear combinations of these variables one can rewrite so 
that all of the equations take the form of `transport equations' in 
the~$n^a\pm s^a$ directions, but with transverse derivatives appearing 
as sources. Particularly relevant are the combinations~$\chi\pm K_{qq}$, 
as they reveal the Raychaudhuri equations. Up to this trivial change of 
variables and the split into time-space derivatives, rather than the 
double-null choice, this is the same system presented in~Ch.~$3$ 
of~\cite{KlaNic03}, where it was also noted that this system is not 
hyperbolic.

\section{Coordinate switched first order generalized harmonic 
gauge}
\label{section:_3+1_GHG11}

\subsection{Double-null Jacobians}\label{subsection:DN_J}

\paragraph*{Time and Radial coordinates:} Let us now abandon 
the idea of evolving in double-null coordinates directly, and 
instead examine how the spacetime could be constructed in these 
coordinates a posteriori, having constructed the spacetime locally 
in the harmonic gauge, for example. This is similar to the strategy 
employed in~\cite{Chr08}. It has also been used in numerically, 
in for example~\cite{GarGunHil07}. Let us work in lower case 
coordinates~$x^{\mu}$. We would like the upper case coordinates to 
satisfy the double-null conditions. As elsewhere,
\begin{align}
N^a=-A\,\nabla^aT\,,\quad\quad S^a= L\,\nabla^aR\,.
\end{align}
with~$X^{\ul{\mu}}=(T,R,\Theta^{\ul{A}})$. We choose~$A=L$ to impose the
double-null gauge, regardless of the angular coordinates. Under 
this condition we define ingoing and outgoing null 
vectors~$L^a,K^a$,
\begin{align}
-N^a+S^a=L^a=L\,\hat{L}^a\,,\quad -N^a-S^a=K^a=L\,\hat{K}^a\,.
\end{align}
The renormalized vectors~$\hat{L}^a$ and~$\hat{K}^a$ generate null 
geodesics in the~$L^a$ and~$K^a$ directions. It is natural 
to define two lower case spatial vectors~$v_{\pm}^a$ and~$s_{\pm}^a$
from the Jacobian according to,
\begin{align}
v^\pm_i &=E_{\pm}\,s^\pm_i = \phi^{\ul{R}}{}_i \mp L^{-1}Wv_i\,.
\end{align}
The vectors~$s_{\pm}^a$ have unit magnitude. Indices~$s_{\pm}$ stand 
for contraction with these vectors. The scalars~$E_{\pm}$ are 
the energy of the ingoing and outgoing congruences as measured by the 
Eulerian observers~$n^a$. In terms of the Jacobian they are,
\begin{align}
E_{\pm} = L^{-1}W \mp \pi^{\ul{R}} \,.
\end{align}
The null geodesic vectors are then,
\begin{align}
\hat{K}^a=-E_-\big(n^a+s_-^a\big)\,,\quad\quad 
\hat{L}^a=-E_+\big(n^a-s_+^a\big)\,.
\end{align}
Straightforward computation then reveals evolution equations,
\begin{align}
\p_t\ln E_{\pm}&= \Lie_{(\beta\pm\alpha s_{\pm})}\ln E_{\pm}
+\alpha\left(K_{s_{\pm}s_{\pm}}\pm\Lie_{s_{\pm}}\ln\alpha\right)\,,
\nonumber\\
\p_tv^{\pm}_i&=\pm\alpha\,D_{(s_{\pm})}v^{\pm}_i\pm E_{\pm}\,D_i\alpha
+\Lie_\beta v^{\pm}_i\,.\label{eqn:DN_Jac}
\end{align}
The hypersurface constraints were used freely to arrive
at this result. Interestingly the term appearing as a source 
in the first equation is essentially a characteristic variable of 
the (first order in time, second order in space) GHG formulation. 
Notice furthermore that, after adjusting the normalization of~$v_{\pm}^i$, 
these can be compared with the results of~\cite{VinGouNov12}, and 
are of course compatible. Note also that the equations for~$E_{\pm}$ 
follow from~$\gamma^{ij}v^{\pm}_iv^{\pm}_j=E^2_{\pm}$. The 
equations~\eqref{eqn:DN_Jac} form a symmetric hyperbolic system 
in~$(E_{\pm},v^{\pm}_i)$ which is equivalent to a system in the 
scalar components~$L^{-1}W$ and~$\pi^{\ul{R}}$ and the vector 
parts~$\phi^{\ul{R}}{}_i$ and~$L^{-1}Wv_i$ of the Jacobian.

\paragraph*{Angular coordinates:} To complete the equations of 
motion for the Jacobian we require a choice 
for~$\phi^{{\ul{A}}}{}_i$ and~$\pi^{\ul{A}}$. We have already seen 
that from the pure gauge point of view the choice~`$\beta^A=b^A$' 
is problematic. To understand how this issue appears in the Jacobian 
formulation, let us assume that the component~$\pi^{\ul{A}}$ takes the 
form,
\begin{align}
\pi^{\ul{A}}= - m^i\phi^{\ul{A}}{}_i\,,\label{eqn:Chr_Ang}
\end{align}
for some known lower case spatial vector~$m^i$. This corresponds 
to determining the angular coordinates by Lie-dragging so 
that~$(n^a+m^a)\nabla_a\Theta^{\ul{A}}=0$. This family 
includes~`$\beta^A=\pm b^A$' by choosing~$m^a=\pm s^a_{\pm}$. Plugging 
the relation~\eqref{eqn:Chr_Ang} into the Jacobian equation of 
motion~\eqref{eqn:Jac_dot} and using the hypersurface constraints 
gives the compact expression,
\begin{align}
\p_t\phi^{{\ul{A}}}{}_i&=\Lie_{(\beta - \alpha\,m)}\phi^{{\ul{A}}}{}_i\,.
\end{align}
If the vector~$m^a$ is given a priori this equation is hyperbolic.
On the other hand if we wish to make~$m^a$ dependent on the other 
components of the Jacobian, the derivative of~$m^a$ in the 
Lie-derivative is problematic, as it is a one-way coupling, in the 
sense that the equations of motion for the angular coordinates 
explicitly depend upon the~$(T,R)$ coordinates in the principal part,
but not vice-versa. This type of coupling is dangerous because it 
can leave non-trivial Jordan blocks in the principal symbol if the 
speeds associated with the~$T,R$ and~$\Theta^{\ul{A}}$ blocks clash. 
Fortunately this discussion suggests two alternatives. The first is 
to construct the angular coordinates using just vectors associated 
with lower case coordinates. For example we could 
choose~$n^a\nabla_a\Theta^{\ul{A}}=0$, or 
even~$t^a\nabla_a\Theta^{\ul{A}}=0$, in which case the angular 
coordinates could correspond to those built directly from the 
lower case coordinates. One possibility would be to use a reference 
metric to build a first order GHG formulation directly in 
spherical-polar coordinates. Then the upper case angular coordinates 
could be exactly those of the lower case system. The second option 
is to make sure that the speeds of the two subsystems do not 
coincide, or that if they do the principal symbol is nevertheless 
diagonalizable. For this we might 
try~$N^a\nabla_a\Theta^{\ul{A}}=W(n^a+v^a)\nabla_a\Theta^{\ul{A}}=0$,
or in other words~$m^a=v^a$, which is also equivalent 
to~$B^{\ul{A}}=0$ when working in upper case coordinates. Since,
\begin{align}
v^i=(E_++E_-)^{-1}\big(v_-^i-v_+^i\big)\,,
\end{align}
we arrive at,
\begin{align}
\p_t\phi^{{\ul{A}}}{}_i&=-\alpha\,v^jD_j\phi^{{\ul{A}}}{}_i
+\tfrac{1}{2}\alpha\,LW^{-1}\phi^{{\ul{A}}}{}_jD_i(v_+^j-v_-^j)
\nonumber\\
&\quad
-\tfrac{1}{2}\alpha\,LW^{-1}\,\pi^{{\ul{A}}}D_i(E_++E_-)
+\pi^{{\ul{A}}}D_i\alpha
+\Lie_{\beta}\phi^{{\ul{A}}}{}_i\,,\label{eqn:m_i=v_i}
\end{align}
after expanding the Lie-derivative.

\paragraph*{Hyperbolicity of the Double-Null Jacobians:} The 
double-null Jacobian system with~$m^i$ given a priori is 
trivially symmetric hyperbolic. Choosing instead~$m^i=v^i$ as 
in~\eqref{eqn:m_i=v_i}, the principal symbol for the 
subsystem with the 
components~$(E_{\pm},v^{\pm}_s,v^{\pm}_A,\phi^{\ul{A}}{}_s)$ is,
\begin{align}
\mathbf{P}^s&=\alpha\,\mathbf{A}^s+\beta^s\,\mathbf{1}\,,
\end{align}
with,
\begin{align}
\mathbf{A}^s&=\left(
\begin{array}{cccc}
 \pm s_{\pm}^s & 0 & 0 & 0  \\
 0 & \pm s_{\pm}^s & 0 & 0  \\
 0 & 0 & \pm s_{\pm}^s & 0  \\
- \tfrac{L}{2\,W}\pi^{\ul{A}} & \pm\tfrac{L}{2\,W}\phi^{\ul{A}}{}_s & 
\pm\tfrac{L}{2\,W}\phi^{\ul{A}}{}_Bq^{AB} & -v^s 
\end{array}\right)\,,
\end{align}
with an index~`$s$' denoting contraction with an arbitrary 
lower case unit spatial vector~$s^i$ and~$q_{ab}=\gamma_{ab}-s_as_b$
as elsewhere. The remaining block of the full principal symbol, 
associated with~$\phi^{\ul{A}}{}_A$, is decoupled, and has no affect on 
the following discussion. The eigenvalues of the principal symbol 
are~$\beta^s\pm \alpha s^s_{\pm}$ and~$\beta^s- \alpha v^s_{\pm}$. For 
strong hyperbolicity we need that the symbol is diagonalizable for 
every~$s^i$. Choose for example~$s^i$ such that,
\begin{align}
(v^i-s^i_-)s_i=0,
\end{align}
then the~`$s_-^s$' and~`$v^s$' eigenvalues coincide, and the principal 
symbol is missing eigenvectors. Therefore the system is again only 
weakly hyperbolic. This type of degeneracy occurs if~$m^i$ is taken to 
be any vector constructed from~$v^i_\pm$, thus we must abandon the 
second alternative suggested by the above discussion. This leaves the 
option to fix~$m^i$ a priori, or to choose a gauge of a different form 
for the angular coordinates, such as the generalized 
harmonic option~\eqref{eqn:Jac_dot_GHG}. At least for the 
choice~`$\beta^A=b^A$' this result was expected, because the pure 
gauge subsystem we examined before is closely related to the system 
for the Jacobians.

\subsection{First order generalized harmonic gauge}
\label{subsection:_3+1_GHG11}

There is much work about formulations of general relativity in 
first order form. For a review, see~\cite{SarTig12}. Of particular 
interest in recent years has been the first order reduction of 
the GHG formulation~\cite{LinSchKid05,Alv02}. 
To a large extent this interest was driven by use inside the 
SpEC numerical relativity code. Here we summarize just the relevant 
features. The first order GHG system is a quasilinear, first 
order symmetric hyperbolic system of the form,
\begin{align} 
\p_t\mathbf{u}&=\mathbf{A}^p\p_p\mathbf{u}+\mathbf{S}\,.
\end{align}
The formulation has a set of constraints compatible with the 
evolution equations. The key feature of the system is that the 
principal part takes the form of a first order reduction of the 
wave equation. This is attained by carefully coupling the 
coordinate choice~$\Box x^{\alpha}=H^{\alpha}$ to the field 
equations~\cite{Bru52} and then reducing to first order. 
Gravitational radiation controlling, constraint preserving 
boundary conditions for the system have been studied and 
implemented~\cite{Rin06,RuiRinSar07,RinLinSch07}. The evolution 
system is now used frequently in the evolution of compact binary 
spacetimes. Of crucial importance in the present work is that 
in the first order reduction, all first derivatives of the 
metric can be expressed in terms of evolved variables without 
taking any derivatives. So, for example if we were to evolve the 
double-null Jacobians~\eqref{eqn:DN_Jac} alongside the GHG 
variables, we can formulate the system so that no coupling occurs 
through derivatives. In other words the Jacobians are minimally 
coupled to the GHG formulation.

\subsection{Coordinate switch and hyperbolicity}
\label{subsection:Change_Coord}

\paragraph*{Coordinate switch:} Suppose, without loss of generality, 
that we are given a first order quasilinear system of the form, 
\begin{align}
\p_T\mathbf{u}&=\left(\,A\,\mathbf{A}^{\ul{p}}+B^{\ul{p}}\,\mathbf{1}
\,\right)\p_{\ul{p}}\mathbf{u}+A\,\mathbf{S}\,.
\end{align}
Now consider the effect of a change 
of independent coordinates on the whole system, so that rather 
than using the upper case~$X^{\ul{\mu}}$ coordinates we employ the 
lower case~$x^\mu$ ones. In the new coordinates the system of 
course retains the same functional form, but now with,
\begin{align}
\big(\mathbf{1}+\mathbf{A}^{\ul{V}}\big)\p_t\mathbf{u}&=
\alpha\,W^{-1}\left(\mathbf{A}^{\ul{p}}\,(\varphi^{-1})^p{}_{\ul{p}}-
\big(\mathbf{1}+\mathbf{A}^{\ul{V}}\big)\Pi^p\right)\p_p\mathbf{u}\nonumber\\
&\quad+\alpha\,W^{-1}\,\mathbf{S}\,.\label{eqn:Change_Coord}
\end{align}
Recall here the various components of the inverse Jacobian
defined by~\eqref{eqn:J_inv} and~\eqref{eqn:J_proj}, and the 
shorthand~$\mathbf{A}^{\ul{V}}=\mathbf{A}^{\ul{i}}\,V_{\ul{i}}$.
It is easily confirmed that any closed constraint subsystem 
remains closed.

\paragraph*{Symmetric hyperbolicity:} The original system is assumed 
to be symmetric hyperbolic, that is, there exists a symmetric positive 
definite symmetrizer~$\mathbf{H}$ such that
\begin{align}
\mathbf{H}\,\left(\,A\,\mathbf{A}^{\ul{p}}+B^{\ul{p}}\,\mathbf{1}
\,\right)\,,\nonumber
\end{align}
is symmetric for each~$\ul{p}$. In the present context, this system 
can be taken as a first order generalized harmonic formulation coupled, 
minimally, to the double-null Jacobians~\eqref{eqn:DN_Jac}. Under 
a smallness assumption on~$V_{\ul{i}}$ symmetric hyperbolicity is 
unaffected by the change of coordinates. Taking the system in the 
form~\eqref{eqn:Change_Coord}, the symmetrizer is unaffected by 
the change of coordinates. If we insist on multiplying on the left 
by the inverse of~$\big(\mathbf{1}+\mathbf{A}^{\ul{V}}\big)$, then 
we pick up exactly a factor of this matrix on the right of the 
modified symmetrizer. But to obtain energy estimates 
for metric components using norms formed from the lower case 
coordinate basis components of tensors more effort may be 
necessary.

\paragraph*{Strong hyperbolicity:} The principal symbol of the system 
in~$X^{\ul{\mu}}$ is,
\begin{align}
\mathbf{P}_X^{\ul{S}}&=\,A\,\mathbf{A}^{\ul{S}}+B^{\ul{S}}\,\mathbf{1}\,,
\end{align}
where on the right hand side superscript~$\ul{S}$ denotes contraction 
with an arbitrary unit upper case spatial covector~$S_i$. 
Multiplying~\eqref{eqn:Change_Coord} by the inverse 
of~$\big(\mathbf{1}+\mathbf{A}^{\ul{V}}\big)$, the principal symbol 
in~$x^\mu$ can be read off,
\begin{align}
\mathbf{P}_x^s&=\alpha\,W^{-1}\left(
\big(\mathbf{1}+\mathbf{A}^{\ul{V}}\big)^{-1}\mathbf{A}^{\ul{s}}
-\Pi^s\,\mathbf{1}\right)\,. 
\end{align}
On the right hand side superscript~$s$ denotes contraction with an 
arbitrary unit lower case spatial covector~$s_i$, and underlined 
superscript~$\ul{s}$ denotes the same contraction, but pushed 
through the transformation~$(\varphi^{-1})^i{}_{\ul{i}}$. Strong hyperbolicity 
is the requirement that for each~$s_i$ there exists a 
symmetrizer~$\mathbf{H}^s$, uniformly symmetric positive definite in~$s_i$,
such that the product of the symmetrizer with the principal symbol is 
symmetric. For quasilinear problems strong hyperbolicity, together with 
some smoothness conditions on the symmetrizer are sufficient to guarantee 
local well-posedness of the initial value problem. These smoothness conditions 
are sometimes included in the definition of strong hyperbolicity. The existence 
of a symmetrizer for fixed~$s_i$ is equivalent to the requirement that the 
principal symbol has a complete set of eigenvectors with real eigenvalues. 
Assuming that the system is strongly hyperbolic in the upper case 
coordinates~$X^{\ul{\mu}}$ this necessary condition can be seen to hold in 
lower case coordinates~$x^\mu$ as follows. This discussion is adapted 
from~\cite{Met13}. Fix~$s_i$. The characteristic polynomial 
of~$\mathbf{P}_X^{\ul{S}}$ has real eigenvalues, and thus we have 
a {\it hyperbolic polynomial} with respect to~$N_{\ul{\mu}}$. It follows that 
if the boost velocity is sufficiently small then~$\mathbf{P}_x^s$ also has 
real eigenvalues~\cite{Hor05}. Estimates of the range over which this 
condition is satisfied can be given. In the presence of a simple block 
structure of~$\mathbf{A}^{\ul{s}}$, the eigenvalues will simply be multiplied 
by those of~$(\mathbf{1}+\mathbf{A}^{\ul{V}})$. Take one of these 
eigenvalues~$\lambda$. Strong hyperbolicity in~$X^{\ul{\mu}}$ implies the 
existence of symmetric positive definite~$\mathbf{H}$ with,
\begin{align}
\mathbf{H}\,\mathbf{L}\equiv
\mathbf{H}\,\Big( \mathbf{A}^{\ul{s}} - (\mathbf{1}+\mathbf{A}^{\ul{V}})
\,(\lambda+\Pi^s)\Big)\,,
\end{align}
symmetric. Suppose that~$\mathbf{u}$ is a non-vanishing vector in the 
nullspace of~$\mathbf{L}$, and thus either an eigenvector or generalized 
eigenvector of~$\mathbf{P}_x^s$. Suppose that it is a generalized 
eigenvector, so that~$\mathbf{u}=\mathbf{L}\mathbf{v}$ for 
some~$\mathbf{v}$. Then,
\begin{align}
\mathbf{u}^T\mathbf{H}\mathbf{u}=\mathbf{u}^T\mathbf{H}\,\mathbf{L}
\mathbf{v}=
\mathbf{v}^T\mathbf{H}\,\mathbf{L}
\mathbf{u}=0\,.
\end{align}
The second equality holds by symmetry of~$\mathbf{H}\,\mathbf{L}$ and 
the third because~$\mathbf{L}\mathbf{u}=0$. Since~$\mathbf{H}$ is symmetric 
positive definite we then have that~$\mathbf{u}=0$. Therefore 
if~$\mathbf{u}$ is in the nullspace of~$\mathbf{L}$ it is a true eigenvector 
of~$\mathbf{P}_x^s$. Thus~$\mathbf{P}_x^s $ has a complete set of 
eigenvectors. Details concerning the remaining uniformity condition 
in~$s_i$ can be found in~\cite{Met13}.

\paragraph*{Discussion:} The advantage of the dual foliation is 
now clear. In section~\ref{subsection:DN} we were unable to 
find a hyperbolic formulation using the double-null coordinates 
directly. Using the dual foliation however the construction of 
such a formulation is essentially trivial. Since the whole 
system is symmetric hyperbolic we do not lose regularity when 
mapping between the two coordinate systems. Regularity will 
however need to be looked at carefully if we are to use the 
double-null radial coordinate all the way to the origin, but 
this issue is only that of using spherical polar coordinates, 
and not related to the dual foliation strategy. With a little 
more book-keeping we can instead work from a standard first 
order in time, second order in space~\cite{HilRic15,GunGar05} 
formulation of GR and avoid the first order reduction, but postpone 
any presentation thereof. One expects that this formulation 
could be treated according to~\cite{CabChrWaf14} to give an economical 
demonstration of a `neighborhood theorem' for the GR characteristic 
initial value problem. {\it Physically} what has happened is that the 
coordinate and gauge degrees of freedom GR were decoupled as much as 
possible. Note that in earlier work a symmetric hyperbolic formulation 
using a double-null gauge was constructed using frame variables coupled 
to the Bianchi equations~\cite{Len01}. In contrast here the aim was 
to arrive at such a formulation with as few changes as possible to
standard formulations used in numerical work. From this practical 
point of view there is always the danger that the 
matrix~$\big(\mathbf{1}+\mathbf{A}^{\ul{V}}\big)$ becomes singular. We 
can mitigate against this by simply evolving for a short but finite 
coordinate time, and then resetting the Jacobian to the identity by 
transforming all the fields into the upper case tensor basis. For 
sufficiently regular data hyperbolicity guarantees short time 
existence so that this procedure can be performed iteratively. Coupled 
with strong theorems on continuation of solutions, this strategy could 
perhaps even be used to guarantee that numerical calculations progress 
successfully into extreme regions of spacetime, although the double-null 
gauge might have to be replaced with some other suitable choice.
For the numerical relativist, probably the simplest summary of 
the double-null Jacobian result is that it is the generalization of the 
dual coordinate frames approach~\cite{SchPfeLin06} employed in the 
SpEC code~\cite{SpEC} to the situation in which two differing time 
coordinates are considered, and where the Jacobians satisfy dynamical 
equations rather than arising as derivatives of algebraic relationships 
between the coordinates.

\section{Conclusion}
\label{section:Conclusion}

Making a dual foliation approach, we have shown that it is possible 
to effectively decouple the choice of coordinates from local 
well-posedness of the field equations of general relativity. 
This was done by evolving a first order reduction of the generalized 
harmonic formulation alongside Jacobians mapping from the desired 
coordinate system to the generalized harmonic one. The important 
example of the double-null gauge was considered. But in fact the 
set of coordinate choices resulting in equations of motion for 
Jacobians that are minimally coupled to the remaining field equations 
is {\it extremely large}, and local well-posedness inside this 
class follows from well-posedness of the coordinate choice alone. It 
is thus expected that with due care, this observation will allow us 
to evolve the generalized harmonic formulation using say, the Maximal 
slicing, and quasi-isotropic spatial coordinates, for example. 
We hope that this would allow for direct comparison of the results 
of~\cite{AbrEva93} in a modern numerical relativity code with minimal 
changes. See~\cite{Sor10,HilBauWey13,HeaLag14,SchRin14,AkbCho15} for 
recent work on the problem. It is important to realize that the role 
of the generalized harmonic formulation in the construction is completely 
auxiliary. In fact any well-behaved formulation of GR is amenable to 
the same trick, perhaps after a suitable reduction to first order, if 
we wish to avoid complicated book-keeping. The generalized harmonic 
formulation is simply the most convenient example. 

For numerical applications it will be necessary to translate the 
constraint preserving boundary conditions~\cite{Rin06a} of the 
harmonic system into the new set of variables and coordinates, but 
no major difficulty is expected in doing so. The obvious next step 
is to systematically implement and test the approach, preferably in 
a simple context. Afterwards we expect to use the double-null 
coordinates to study the critical collapse of gravitational waves
using the~\texttt{bamps} code~\cite{Bru11,HilWeyBru15,BugDieBer15}. 
\\ \\

\acknowledgments

I am grateful to Bernd Br\"ugmann, Sascha Husa, Ronny Richter, 
Andreas Weyhausen and especially Juan Antonio Valiente-Kroon for 
helpful discussions. Thanks go also to Tim Dietrich, David 
Garfinkle and Milton Ruiz for their feedback on the manuscript. 
Finally I wish to thank the entire UIB relativity group for 
their warm hospitality during a recent visit in which part of 
the work was completed. Many of the calculations presented here 
were performed using {\tt mathematica} with the package {\tt xTensor} 
by Jos\'e-Mar\'ia Mart\'in-Garc\'ia~\cite{xAct_web}. The notebooks 
are available at the website of the author~\cite{HilWebsite}. This 
work was supported in part by DFG grant SFB/Transregio~7 
`Gravitational Wave Astronomy'. 

\appendix

\bibliographystyle{unsrt}
\bibliography{DF}{}

\end{document}